\documentclass[twocolumn]{webofc}
\usepackage[varg]{txfonts}   % Web of Conferences font
\usepackage{bm,ulem}
\usepackage{graphicx, hyperref}
\usepackage{color}

\begin{document}
%%%%%%%%%%%%%%%%%%%%%%%%%%%
\title{Discontinuous shear thickening of a moderately dense inertial suspension of hydrodynamically interacting frictionless soft particles}
\author{
    \firstname{Satoshi} \lastname{Takada}\inst{1}\fnsep
    \thanks{\email{takada@go.tuat.ac.jp}}
    \and
    \firstname{Kazuhiro} \lastname{Hara}\inst{2}\fnsep
    \thanks{Present address: 
    Suzuki Motor Corporation, 
    300 Takatsukacho, Minami-ku, Hamamatsu, Shizuoka 432-8611, Japan}
    \and
    \firstname{Hisao} \lastname{Hayakawa}\inst{3}\fnsep
    \thanks{\email{hisao@yukawa.kyoto-u.ac.jp}}
}

\institute{
    Department of Mechanical Systems Engineering and Institute of Engineering, 
    Tokyo University of Agriculture and Technology, 
    2-24-16 Naka-cho, Koganei, Tokyo 184-8588, Japan
    \and
    Department of Industrial Technology and Innovation,
    Tokyo University of Agriculture and Technology,
    2-24-16 Naka-cho, Koganei, Tokyo 184--8588, Japan
    \and
    Yukawa Institute for Theoretical Physics, 
    Kyoto University,
    Kitashirakawa Oiwakecho, Sakyo-ku, Kyoto 606-8502, Japan
}

%%%%%%%%%%%%%%%%%%%%%%%%%%%
\abstract{
We demonstrate that discontinuous shear thickening (DST) can occur even in moderately dense, inertial suspensions of hydrodynamically interacting, frictionless soft particles.
Using the Lubrication-Friction Discrete Element Method, our simulations reveal that DST can emerge at lower particle densities, provided that both the inertia of the suspended particles and their softness are sufficiently pronounced.
Furthermore, we show that, under these conditions, the DST behavior obtained from the simulation qualitatively agrees with that predicted by kinetic theory, even without accounting for hydrodynamic interactions.
These findings expand the understanding of DST in soft particle systems and highlight the importance of particle inertia and softness in controlling rheological behavior.
}
\maketitle

%%%%%%%%%%%%%%%%%%%%%%%%%%%

%%%%%%%%%%%%%%%%%%%%%%%%%%%%%%%%%%%%%%%%%%%%%%%%%%
\section{Introduction}
Discontinuous shear thickening (DST) is a phenomenon in which the viscosity of a suspension undergoes an abrupt increase at a critical shear rate~\cite{Wagner09, Brown14, Ness22}.
This sudden change in viscosity is accompanied by a sharp variation in the normal stress difference~\cite{Laun94, Cwalina14}.
DST has been observed in various systems, including suspensions of solid particles in liquid media and frictional dry granular materials~\cite{Otsuki11}, where it is closely associated with the process of shear jamming~\cite{Bi11, Fall15, Otsuki20, Pradipto20}.

Due to its significant implications for industrial applications, DST plays a crucial role in the design and performance of devices such as sporting equipment~\cite{Fischer06}, traction control systems~\cite{Laun91}, protective vests~\cite{Lee03}, and materials handling processes.
Beyond its industrial relevance, DST is of great interest to physicists as a model for studying nonequilibrium phase transitions.
A comprehensive understanding of the physical mechanisms underlying DST is therefore essential not only for advancing theoretical knowledge but also for optimizing industrial applications that rely on densely packed particles.

Despite its importance, the exact physical origin of DST remains a topic of ongoing debate. 
Current theories suggest that the most likely cause of DST is inter-particle frictional forces~\cite{Otsuki11, Wyart14}.
However, alternative mechanisms, such as order-disorder transitions~\cite{Hoffman} and hydrodynamic clusters~\cite{Brady, Bender96}, have also been proposed as contributing factors.

It is well established theoretically that a DST-like phenomenon, driven by an ignited-quenched transition, occurs in inertial suspensions.
This can be regarded as a model of aerosols, in which particle collisions play a key role in this DST-like behavior~\cite{Friedlander, Koch01}.
Furthermore, several theoretical studies have investigated inertial suspensions composed of hard-core, frictionless particles using kinetic theory, which neglects hydrodynamic interactions between particles~\cite{Hayakawa19, Tsao95, Sangani96, Garzo12, Saha, Hayakawa17, Takada20}.
These studies indicate that the transition from the ignited to the quenched phases becomes continuous when the volume fraction ($\varphi$) exceeds a few percent~\cite{Sangani96, Saha, Hayakawa17, Takada20}.
This behavior contrasts with DST commonly observed in colloidal suspensions, where DST typically occurs only in highly dense suspensions.

Recently, Sugimoto and Takada~\cite{Sugimoto20} theoretically found a two-step DST in dilute inertial suspensions of soft, frictionless particles.
The first DST-like change corresponds to an ignited-quenched transition, while the second originates from an exploded-continuous shear thickening (E-CST) transition~\cite{Sugimoto20, FullPaper}.
We have provided a complementary analysis based on kinetic theory for moderately dense inertial suspensions of frictionless soft particles~\cite{FullPaper}, showing that this second DST-like change persists even in moderately dense situations.

In this paper, we demonstrate the occurrence of the second DST-like change of the inertial suspensions when hydrodynamic interactions are present.
A key motivation of the present study is to further investigate the role of hydrodynamic interactions among particles in the behavior of DST.
Although previous studies of colloidal suspensions have ignored the effects of inertia in the equation of motion, it is important to clarify the roles of both hydrodynamic interactions and inertia in determining the rheology of the system we are modeling.
For this purpose, we focus on situations in which the fluid motion can be described by the Stokes equation.
However, incorporating the effects of long-range interactions between particles remains a challenging task.
As a first step, we adopt the Lubrication-Friction Discrete Element Method (LF-DEM), which considers only short-range lubrication forces between particles~\cite{Mari15}.

%%%%%%%%%%%%%%%%%%%%%%%%%%%%%%%%%%%%%%%%%%%%%%%%%%
\section{Simulation model}
As mentioned in the previous section, we adopt the LF-DEM~\cite{Mari15}. 
Such a simplification, whereby we ignore long-range hydrodynamic interactions, may be justified in relatively dense suspensions.

We consider $N$ monodisperse frictionless soft particles (each particle of mass $m$ and diameter $d$), suspended in a fluid and confined within a three-dimensional cubic box with linear size $L$.
We assume that the harmonic potential describes the contact force between particles
\begin{equation}\label{harmonic_potential}
    U(r)=\frac{\varepsilon}{2}\left(1-\frac{r}{d}\right)^2\Theta\left(1-\frac{r}{d}\right) ,
\end{equation}
where $r$ is the inter-particle distance, $\varepsilon$ is the energy scale to characterize the repulsive interaction, and $\Theta(x)$, defined as $\Theta(x)=1$ ($x\ge 0$) and $0$ ($x<0$) is the step function.
Although clustering effects caused by attractive interactions between particles cannot be ignored in realistic situations, such effects are suppressed if particles are charged~\cite{Derjaguin41,Israelachvili}.
Moreover, particles are also prevented from clustering if the temperature is sufficiently high~\cite{Hayakawa19}. 

The equation of motion of the suspended particle $i$ (its position $\bm{r}_i$) under simple shear with the shear rate $\dot\gamma$ and the peculiar momentum $\bm{p}_i\equiv m(\bm{v}_i-\dot\gamma y_i\hat{\bm{e}}_x)$ with the unit vector $\hat{\bm{e}}_x$ parallel to the $x$ direction and the velocity $\bm{v}_i$ of the $i-$th particle is given by
\begin{equation}\label{eq:Langevin}
    \frac{d\bm{p}_i}{dt} 
    = -\sum_{j\neq i}\frac{\partial U(r_{ij})}{\partial \bm{r}_{ij}}
    - \sum_j \overleftrightarrow{\zeta_{ij}} \bm{p}_j 
    +\bm{\zeta}^\mathrm{Sh}_i
    + \bm{\xi}_i,
\end{equation}
with $\bm{r}_{ij}\equiv \bm{r}_i - \bm{r}_j$ and $r_{ij}\equiv |\bm{r}_{ij}|$. 
The noise term $\bm{\xi}_i(t)=\xi_{i,\alpha}(t)\hat{\bm{e}}_\alpha$ satisfies the fluctuation–dissipation relation,
\begin{equation}
    \langle \bm{\xi}_i(t) \rangle = \bm{0},\quad
    \langle \xi_{i,\alpha}(t) \xi_{j,\beta}(t^\prime)\rangle
    = 2m \zeta T_\mathrm{env} \delta_{ij}\delta_{\alpha\beta} \delta(t-t^\prime),
\end{equation}
where $\langle \cdot \rangle$ expresses the average over the noise, $\zeta$ is the drag coefficient proportional to the solvent viscosity, and $T_\mathrm{env}$ is the environmental temperature.
It is noted that the Boltzmann constant is set to be unity in this paper.
The tensor $\overleftrightarrow{\zeta_{ij}}$ is the resistance matrix in the Stokes flow with the existence of dimples given by
\begin{equation}
    \zeta_{ij,\alpha\beta}
    = 
    \begin{cases}
    \displaystyle 3\pi \frac{\eta_0 d_\mathrm{H}}{m}\delta_{\alpha\beta}
    +\displaystyle \sum_{k\neq i} \frac{1}{m}A_{ik,\alpha\beta}^{(11)}
    \Theta(d-r_{ik}) & (i=j)\\
    \displaystyle -\frac{1}{m}A_{ij,\alpha\beta}^{(11)}
    \Theta(d-r_{ij}) & (i\neq j)
    \end{cases},
\end{equation}
where the explicit expression of the coefficient $A_{ij,\alpha\beta}^{(11)}$ is given in Refs.~\cite{FullPaper, KimKarrila}, $\eta_0(=\zeta/(3\pi d))$ is the viscosity of the solvent, and $d_\mathrm{H}\equiv d/(1+\delta)$ with the roughness parameter $\delta$~\cite{Mari15,Pradipto20} is the cutoff length of the lubrication force with the hydrodynamic diameter.
Although smooth hard-core particles are not allowed to contact each other~\cite{HappelBrenner,Jeffrey84, Jeffrey92, KimKarrila, Ichiki13}, the LF-DEM allows contact between rough particles thanks to the roughness parameter $\delta$.
This parameter equates to a simplified description of dimples on the surface of particles.
We also note that $\bm{\zeta}^\mathrm{Sh}_i$ is the contribution from the shear flow due to the lubrication force given by
\begin{equation}
    \zeta^\mathrm{Sh}_{i,\alpha}
    = 2\eta_0 \dot\gamma
    \sum_j \widetilde{G}^{(11)}_{ij,xy\alpha},
\end{equation}
where the explicit expression of the coefficient $\widetilde{G}^{(11)}_{ij,xy\alpha}$ is given in Refs.~\cite{FullPaper, KimKarrila}.
The terms containing $\overleftrightarrow{\zeta_{ij}}$ and $\bm{\zeta}^{\rm Sh}$ in Eq.~\eqref{eq:Langevin} originate from the hydrodynamic force acting on the $i$--th particle.

In this system, the stress tensor is given by 
\begin{align}
    \sigma_{\alpha\beta}
    &= \frac{1}{NL^3}\sum_i \Bigg\langle \left(-m v_{i,\alpha}v_{i,\beta} 
    + \frac{1}{2}\sum_{j\neq i}r_{ij,\alpha}\frac{\partial U(r_{ij})}{\partial r_{ij, \beta}}\right.\nonumber\\
    &\hspace{6em}\left.
    + \frac{1}{2}\sum_{j\neq i}\sigma_{ij,\alpha\beta}^\mathrm{H}\right) \Bigg\rangle + \frac{1}{L^3}\sigma^\mathrm{St}_{\alpha\beta},
    \label{eq:sigma_hydro}
\end{align}
where $L^3$ is the volume of the system.
Here, $\sigma^\mathrm{St}_{\alpha\beta}$ is the stress from the Stokes flow given by~\cite{KimKarrila}
\begin{equation}
    \sigma^\mathrm{St}_{\alpha\beta}
    = \frac{5}{12}\pi d_\mathrm{H}^3 \eta_0\dot\gamma
    (\delta_{\alpha x}\delta_{\beta y} + \delta_{\alpha y}\delta_{\beta x}),
\end{equation}
and $\sigma^\mathrm{H}_{ij, \alpha\beta}$ is the hydrodynamic stress between the $i$--th and the $j$--th particles given by~\cite{KimKarrila}
\begin{equation}
    \sigma^\mathrm{H}_{ij, \alpha\beta}
    = -2\eta_0 G^{(11)}_{ij,\alpha\beta\gamma}V_{ij,\gamma}
    + 2\eta_0 \dot\gamma M^{(1)}_{ij,\alpha\beta} ,
\end{equation}
with $\bm{V}_i\equiv \bm{v}_i - \dot\gamma y_i \hat{\bm{e}}_x$. 
The explicit expressions of $G^{(11)}_{ij,\alpha\beta\gamma}$ and $M^{(1)}_{ij,\alpha\beta}$ are presented in Ref.~\cite{KimKarrila}.

Let us introduce the P\'{e}clet number $\mathrm{Pe}$ defined as~\cite{Hunt02}
\begin{equation}
    \mathrm{Pe}
    \equiv \frac{3\pi \eta_0 d^3}{4T_\mathrm{env}}\dot\gamma.
    \label{eq:Pe}
\end{equation}
Using $\mathrm{Pe}$, we introduce the dimensionless viscosity $\eta^*$ as
\begin{equation}
    \eta^*
    \equiv\frac{\sigma_{xy}}{n T_\mathrm{env}\mathrm{Pe}}.
\end{equation}    
To characterize the system, we also introduce the volume fraction of particles, the magnitude of the environmental temperature $\xi_\mathrm{env}$ and the softness $\varepsilon$, respectively, as~\cite{FullPaper}
\begin{equation}
    \varphi \equiv \frac{1}{L^3}\frac{\pi}{6}Nd^3,\quad
    \varepsilon^*\equiv \frac{\varepsilon}{md^2\zeta^2},\quad
    \xi_\mathrm{env}\equiv \sqrt{\frac{T_\mathrm{env}}{m}}\frac{1}{d\zeta}.
\end{equation}

In the simulation, we adopt SLLOD dynamics~\cite{Evans} with the aid of the Lees-Edwards boundary condition~\cite{Lees72}.
As far as we have checked, the uniform flow is stable once the system reaches a steady state.
We examine $N=1000$ and $10$ ensemble averages in the simulations.
The time increment for the simulation is chosen as $\Delta t=10^{-2}\mathrm{min}(d/\sqrt{2T/m}, d\sqrt{m/\varepsilon})$, where $\mathrm{min}(a, b)$ chooses the smaller one between $a$ and $b$.
We control $\mathrm{Pe}$, $\xi_\mathrm{env}$, and $\varepsilon^*$ in the range $0.25\le \mathrm{Pe}\le 25$, and $10^2\le \varepsilon^*\le 10^8$, respectively, and fix $\delta=0.05$ and $\xi_\mathrm{env}=1.0$ in this paper.

%%%%%%%%%%%%%%%%%%%%%%%%%%%%%%
\section{Rheology}
%%%%%%%%%%%%%%%%%%%%%%%%%%%%%%
%%%%%%%%%%%%%%%%%%%%%%%%%%%%%%
\begin{figure}[htbp]
    \centering
    \includegraphics[width=0.8\linewidth]{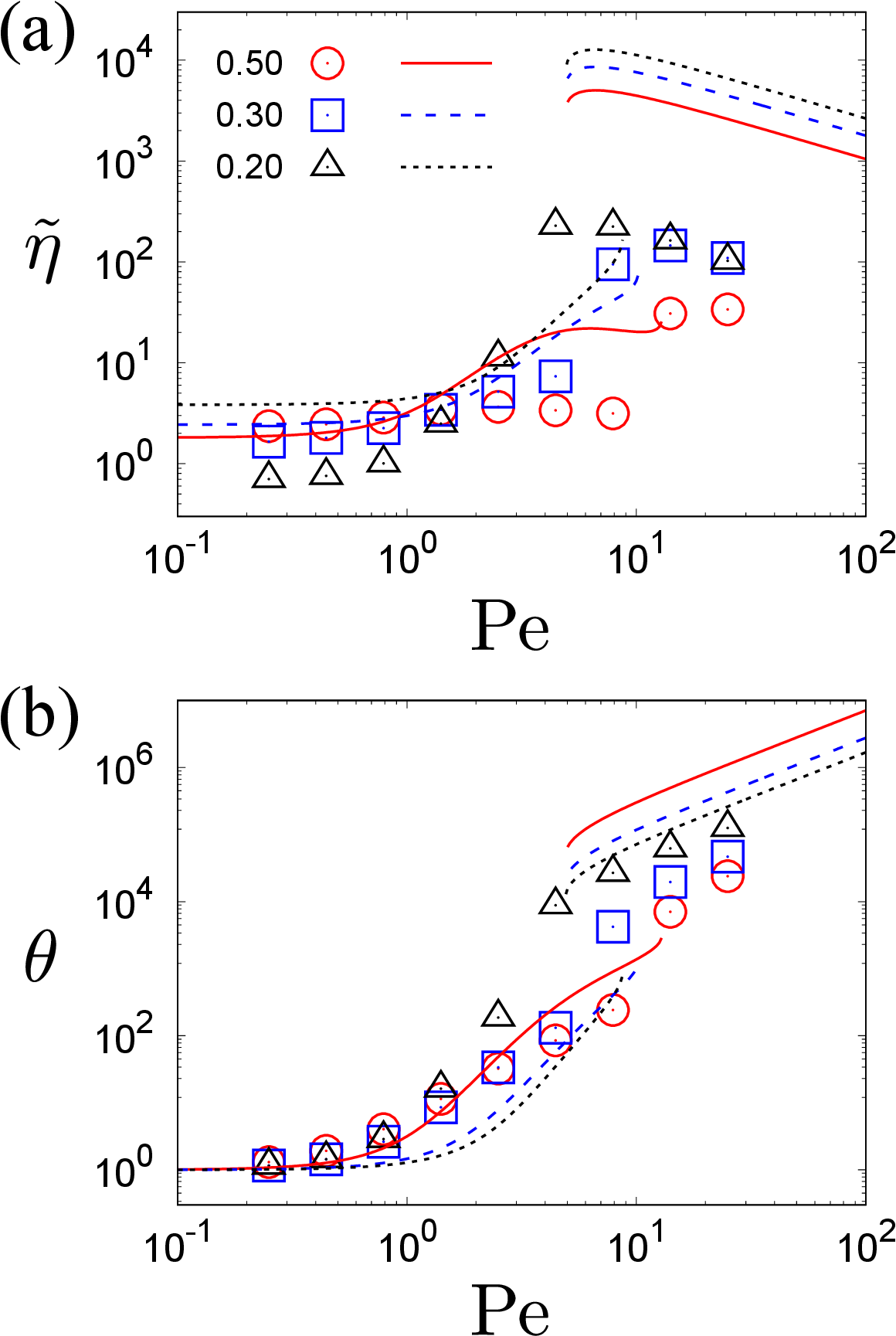}
    \caption{Plots of (a) $\tilde{\eta}$ and (b) $\theta$ against $\mathrm{Pe}$ for $\varphi=0.20$ (open triangles and dotted line), $0.30$ (open squares and dashed line), and $0.50$ (open circles and solid line) with fixed values of $\delta=0.05$, $\varepsilon^*=10^4$, and $\xi_\mathrm{env}=1.0$.
    }
    \label{fig:flow_curves_hydro_0.40}
\end{figure}
%%%%%%%%%%%%%%%%%%%%%%%%%%%%%%

Figure~\ref{fig:flow_curves_hydro_0.40} illustrates the dependence of the scaled viscosity $\tilde{\eta} \equiv \eta^*/\eta_\mathrm{a}$ and the dimensionless temperature $\theta \equiv T/T_\mathrm{env}$ on the P\'{e}clet number $\mathrm{Pe}$ for various volume fractions $\varphi$, with fixed parameters $\delta = 0.05$, $\varepsilon^* = 10^4$, and $\xi_\mathrm{env} = 1.0$. 
Here, the kinetic temperature is defined as $T \equiv \sum_{i=1}^N \langle m_i \bm{V}i^2 \rangle / (3N)$, and the apparent viscosity $\eta_\mathrm{a}$ in the low shear limit is empirically expressed as $\eta_\mathrm{a} = 1 + (5/2)\varphi + 4\varphi^2 + 42\varphi^3$~\cite{deKruif85}.
While the term up to $\mathcal{O}(\varphi^2)$ in $\eta_{\rm a}$ is theoretically established for hard-sphere suspensions~\cite{Batchelor72}, to our knowledge, the expression beyond this order lacks a rigorous theoretical foundation. 
Figure~\ref{fig:flow_curves_hydro_0.40} reveals a striking feature: discontinuous shear thickening (DST) occurs in the range $1 \lesssim \mathrm{Pe} \lesssim 10$.
In the upper (high-stress) branch, the kinetic theory without hydrodynamic interactions~\cite{FullPaper} qualitatively captures the shear-thinning trend, although the quantitative agreement with the hydrodynamic simulation is poor. 
Despite this discrepancy, the results from the LF-DEM qualitatively align with the theoretical framework presented in Ref.~\cite{FullPaper}, where DST is interpreted as an ignited-quenched or E-CST transition in inertial suspensions.
These results confirm that both DST and the associated inertial transitions for soft, frictionless particles persist even when hydrodynamic interactions are incorporated. 
This suggests the robustness of the theoretical framework in capturing the essential features of inertial suspension rheology.

%%%%%%%%%%%%%%%%%%%%%%%%%%%%%%
\begin{figure}[htbp]
    \centering
    \includegraphics[width=0.8\linewidth]{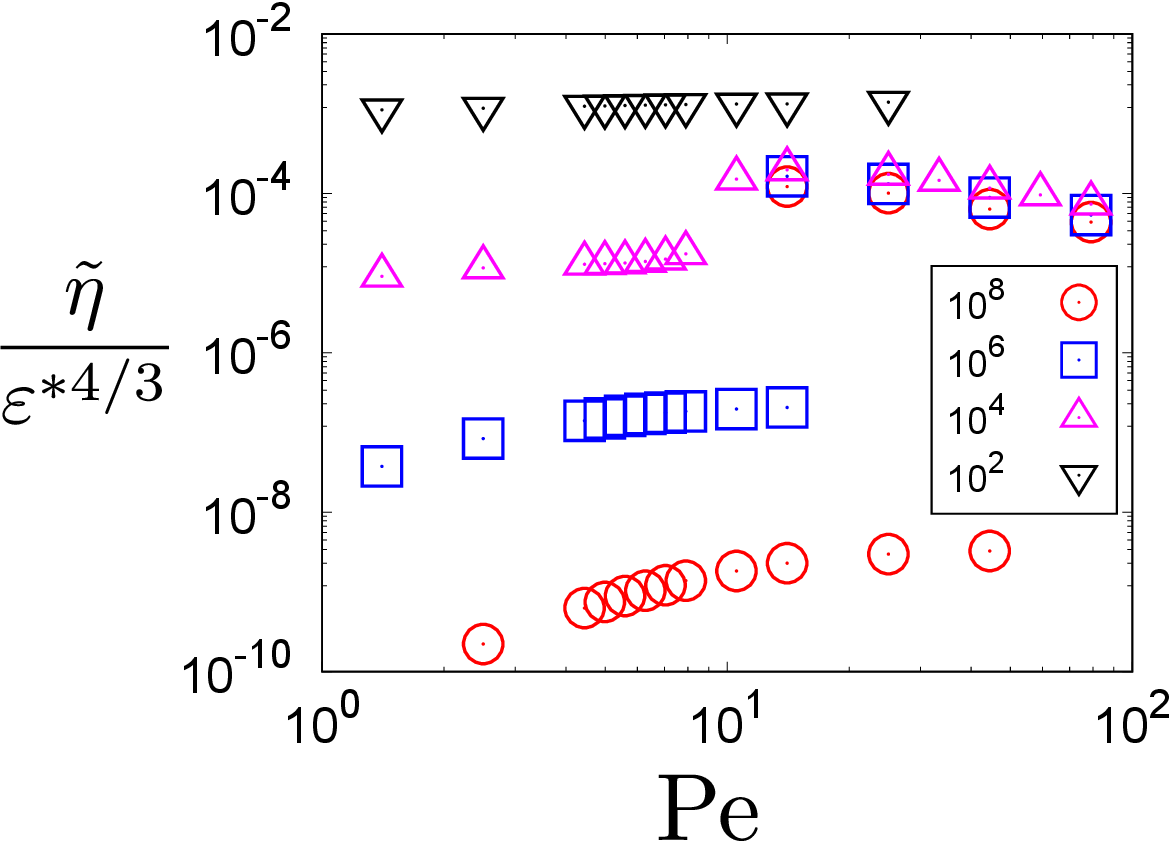}
    \caption{Plots of $\tilde{\eta}/\varepsilon^{*4/3}$ against $\mathrm{Pe}$ for various $\varepsilon^*$ with fixed values of $\varphi=0.40$, $\delta=0.05$, and $\xi_\mathrm{env}=1.0$.
    }
    \label{fig:flow_curves_eps}
\end{figure}
%%%%%%%%%%%%%%%%%%%%%%%%%%%%%%
Figure~\ref{fig:flow_curves_eps} presents the variation of the scaled viscosity $\tilde{\eta}$ as a function of the softness parameter $\varepsilon^*$. 
A clear scaling behavior, $\tilde{\eta} \sim \varepsilon^{*4/3}$, is observed in the exploded phase, except in the regime of extremely soft particles. 
This scaling trend is consistent with the findings reported in Ref.~\cite{FullPaper}.

%%%%%%%%%%%%%%%%%%%%%%%%%%%%%%
\section{Concluding remarks}
In this paper, we have successfully demonstrated the existence of DST-like behavior, characterized by abrupt changes in viscosity and kinetic temperature, in moderately dense inertial suspensions composed of frictionless soft particles using the LF-DEM. 
Our results show the discontinuous changes in both viscosity and kinetic temperature, which qualitatively agree with those by the kinetic theory without accounting for hydrodynamic interactions. 
Additionally, we have identified a novel mechanism for DST, driven by the E-CST transition in frictionless soft particles.

While our model predicts significant and discontinuous changes in both viscosity and kinetic temperature, it is challenging to observe such behavior in liquid suspensions. 
For instance, the drastic increase in kinetic temperature in the ignited or exploded phase could potentially lead to the evaporation of the solvent due to the strong stirring effects of suspended particles, though our model does not include such effects. 
Furthermore, even in gas suspensions, particle melting could pose practical difficulties. 
Nevertheless, the indication of DST-like changes in viscosity and kinetic temperature for frictionless soft particles in moderately dense suspensions is a valuable insight and warrants further exploration in future studies.

A natural extension of this work is to examine the role of interparticle friction, particularly in regimes of large $\eta^*$ that are relevant to typical experimental conditions in colloidal suspensions. 
This will be the subject of our forthcoming study.

Finally, it is important to note that the current simulations include only lubrication interactions. 
A comprehensive understanding of the system requires incorporating long-range hydrodynamic effects, which we leave as a direction for future work.

%%%%%%%%%%%%%%%%%%%%%%%%%%%%%%
\section*{Acknowledgements}
The authors thank Takeshi Kawasaki, Takashi Uneyama, Michio Otsuki, and Pradipto for their helpful comments.
This research was partially supported by Grants-in-Aid from MEXT for Scientific Research (Grants No.~\href{https://kaken.nii.ac.jp/grant/KAKENHI-PROJECT-20K14428}{JP20K14428}, No.~\href{https://kaken.nii.ac.jp/grant/KAKENHI-PROJECT-21H01006}{JP21H01006}, No.~\href{https://kaken.nii.ac.jp/grant/KAKENHI-PROJECT-24K06974}{JP24K06974}, and No.~\href{https://kaken.nii.ac.jp/grant/KAKENHI-PROJECT-24K07193/}{JP24K07193}).
H.H. was also supported by the Kyoto University Foundation. 
The authors thank YITP activity YITP-X-21-13.
%%%%%%%%%%%%%%%%%%%%%%%%%%%%%%

%%%%%%%%%%%%%%%%%%%%%%%%%%%%%%

\end{document}